\documentclass[floats,aps,showpacs,twocolumn]{revtex4}

\usepackage[dvips]{graphicx}
\usepackage{latexsym}
\usepackage{amsfonts}
\usepackage{amssymb}
\usepackage{float}
\usepackage{times}

\newcommand{\dc}{{\cal D}}
\newcommand{\ec}{{\cal E}}
\newcommand{\nc}{{\cal N}}
\newcommand{\rmb}{\rho_{\text {\tiny MB}}}
\newcommand{\sca}{{\cal S}}

\newlength{\figwidth}
\begin{document}

\title{Level Density of a Bose Gas and Extreme Value Statistics}

\author{A. Comtet $^{1,2}$ , P. Leboeuf $^{1}$ and Satya N. Majumdar $^{1}$ }

\affiliation{ $^{1}$ Laboratoire de Physique Th\'eorique et Mod\`eles
  Statistiques, CNRS, Universit\'e de Paris-Sud, UMR 8626, 91405 Orsay Cedex,
  France, $^{2}$ Institut Henri Poincar\'e, 11 rue Pierre et Marie Curie,
  75005 Paris, France}

\date{\today}

\begin{abstract}

  We establish a connection between the level density of a gas of
  non--interacting bosons and the theory of extreme value statistics.
  Depending on the exponent that characterizes the growth of the underlying
  single--particle spectrum, we show that at a given excitation energy the
  limiting distribution function for the number of excited particles follows
  the three universal distribution laws of extreme value statistics,
  namely Gumbel, Weibull and Fr\'echet. Implications of this result, as well
  as general properties of the level density at different energies, are
  discussed.

\end{abstract}


\maketitle

The level density is an essential quantity in determining the thermodynamic
properties of closed quantum systems. In interacting many--body (MB) systems
its computation is in general a difficult problem. The most common framework
is a mean--field approximation, where a gas of independent (quasi-)particles
moves in an average self--consistent potential. In this case, the energy of
the gas is expressed as the sum of the occupied single--particle (SP)
energies. The computation of the MB level density thus reduces to a
combinatorial problem: counting the number of ways into which the energy can
be distributed among the particles. The level density has been extensively
studied in fermionic systems, where detailed experimental data exists at
different excitation energies and quantum numbers (see, for some recent
progresses in this field, Refs.\cite{ld,lmr}). In spite of the experimental
breakthroughs of the 90's and of the many interesting developments that
followed, the case of bosonic systems is much less known. Studies of the
spectral properties have concentrated on the low energy range of the spectrum
of the condensate phase, where collective effects and interactions play a
crucial role.

Our aim here is to compute, within an independent--particle approximation, the
asymptotic properties of the MB level density ${\rmb}(E,N)$ of a Bose gas as a
function of the excitation energy $E$ and the particle number $N$. We will
first consider two extreme regimes that correspond to the degenerate quantum
and the classical limits of the gas. The level density in these two extreme
cases behaves quite differently as a function of energy. In the former case,
where one takes the $N\to \infty$ limit first keeping the energy $E$ finite,
the level density ${\rmb}(E,\infty)$ increases with energy in a
stretched-exponential manner for large $E$. In contrast, in the classical
limit where one keeps $N$ finite and takes the large $E$ limit, the level
density increases with energy in a power-law fashion. This leads to a natural
question: what happens in between these two extreme regimes where both $E$ and
$N$ are large but finite? The main result of this Letter is to show that in
this intermediate regime the level density displays a rich variety of scaling
behaviors depending on the SP spectrum and has an interesting connection to
the extreme value statistics (EVS) of independent random variables.

To explore this intermediate regime, we stay close to the degenerate gas limit
and compute explicitly the effect of a finite number of bosons $N$ on the
level density. In this regime, in a given configuration of excitation energy
$E$, only a fraction of the particles contribute to $E$, the rest remain in
the ground--state. However, the ground--state occupancy and consequently the
number of excited bosons fluctuate among different configurations belonging to
the same excitation energy $E$. These fluctuations may be small or anomalously
large depending on the SP spectrum. To obtain a quantitative estimate of these
fluctuations, we compute explicitly the distribution of the number of excited
particles for a fixed (but large) $E$. We will show that the fraction of
configurations at excitation energy E with $N$ or less excited bosons, among
all possible configurations belonging to the level $E$, has a limiting
distribution (when suitably scaled) for large $N$ and large $E$. Depending on
the index $\nu$ that controls the growth of the SP number of states (cf
Eq.~(\ref{nav}) below), we show that three limiting distributions emerge,
namely Gumbel, Weibull and Fr\'echet distributions. Interestingly,
precisely the same three limiting distributions characterize the EVS of
independent random variables~\cite{Gumbel}, a field that has seen a recent
resurgence of interests~\cite{EVS}.

Our work thus provides a link between these two a priori unrelated fields,
namely the combinatorial problem associated with a non--interacting Bose gas
and the EVS. We believe that this link is of interest in different branches of
physics (such as in the computation of black hole entropy \cite{dabhol}),
mathematics and computer science. For instance, it is well known that the
computation of the level density of a Bose gas in a one--dimensional (1D)
harmonic potential (equidistant SP spectrum) is directly related to the theory
of partitions of an integer \cite{vu,WH}. The theory of partitions has given
rise to deep and fundamental results in mathematics, some of them related to
unique developments by their originality and importance \cite{andrews}. Hence
our results also provide a link between the number partitioning problem and
the EVS, generalizing a theorem of Erd\"os and Lehner \cite{el} (see also
\cite{ak,ht}) which states that the number of summands in a random partition
of an integer is asymptotically distributed with the Gumbel law.

We consider non--interacting bosons confined by some single--particle
potential whose energy levels are $\epsilon_{j}$, $j=0, 1, 2, \ldots$. We set
$\epsilon_0=0$ without loss of generality. Each configuration $\{ n_{j}\} $ of
the gas is characterized by an excitation energy $E= \sum_{j=1}^{\infty} n_{j}
\epsilon_{j}$ and a total number of particles $N = \sum_{j=0}^{\infty} n_{j}$,
where $n_{j}=0, 1, 2, \ldots$ is the occupation number of the $j$-th SP level
in that configuration. The level density at excitation energy $E$ of a gas of
$N$ bosons is given by
\begin{equation} \label{rhodef}
\rmb (E,N) = \sum_{ \{n_j\} } 
\delta\left(E-\sum_{j=1}^{\infty} n_{j} \epsilon_{j}\right) 
\delta \left(N-\sum_{j=0}^{\infty} n_{j}\right) \ .
\end{equation}
The number of excited bosons is simply $N_{\rm ex}=N-n_0=\sum_{j=1}^{\infty}
n_j$. Since $n_0\ge 0$, it follows that $N_{\rm ex}\le N$. Thus, if one just
keeps track of only the excited bosons, it is an easy exercise to show that
$\rmb (E,N)$ in Eq.~(\ref{rhodef}) can alternately be interpreted as the
number of configurations with energy $E$ and with $N_{\rm ex}\le N$. Thus,
when $N \rightarrow \infty$, $\rmb (E,\infty)$ simply counts the total number
of configurations at energy $E$.

A convenient way to express Eq.~(\ref{rhodef}) is by means of an inverse
Laplace transform
\begin{equation} \label{rholap}
\rmb (E,N) = \frac{1}{(2\pi i)^2} \int_{a-i\infty}^{a+i\infty} 
\! \! d \beta
\int_{b-i\infty}^{b+i\infty} 
\! \! d \alpha \ {\rm e}^{\sca (\alpha,\beta)} \ ,
\end{equation}
where
\begin{equation} \label{sdef}
\sca (\alpha,\beta) = -\beta \Omega (\alpha,\beta) + \beta E - \alpha N 
\end{equation}
is the entropy,
\begin{equation} \label{omegadef}
\Omega (\alpha, \beta) = - \int d \epsilon 
\frac{\nc (\epsilon)}{{\rm e}^{\beta \epsilon - \alpha}-1}
\end{equation}
 the grand potential of the gas, and
\begin{equation} \label{ndef}
\nc (\epsilon)=\int^\varepsilon \rho(\varepsilon) d\varepsilon
\end{equation}
the integrated density of states expressed in terms of the SP density of
states $\rho (\epsilon)= \sum_j \delta (\epsilon - \epsilon_j)$. In
Eq.~(\ref{rholap}), $a$ and $b$ are real parameters such that all the poles of
the integrand are to the left of the integration path.

A saddle point approximation with respect to the auxiliary parameters $\alpha$
and $\beta$ of the integrals in Eq.~(\ref{rholap}) yields \cite{fow}
\begin{equation} \label{rhosp}
\rmb (E,N) = {\rm e}^{\sca (\alpha,\beta)}/ \ 2\pi \sqrt{|\dc (\alpha,\beta)|} \ 
\end{equation}
where $\cal D(\alpha,\beta)$ is the determinant of the second derivatives of
$S(\alpha,\beta)$. The dependence on $N$ and $E$ in Eq.~(\ref{rhosp}) arises
from the saddle point conditions that determine implicitly the values of
$\alpha$ and $\beta$ in terms of $N$ and $E$
\begin{eqnarray} \label{spc1} 
  \nc (\alpha, \beta) &=& \int d \epsilon
  \frac{\rho (\epsilon)}{{\rm e}^{\beta \epsilon - \alpha}-1} = N \ , \\
  \ec (\alpha, \beta) &=& \int d \epsilon \frac{\epsilon \ \rho
  (\epsilon)}{{\rm e}^{\beta \epsilon - \alpha}-1} = E \ , \label{spc2}
\end{eqnarray}
where $\nc (\alpha, \beta)$ and $\ec (\alpha, \beta)$ are the particle number
and energy functions of the gas, respectively. We will work here in the
leading order approximation $\rmb (E,N) \approx {\rm e}^{\sca
(\alpha,\beta)}$, and thus ignore the prefactor in Eq.~(\ref{rhosp}).

In Eqs.(\ref{spc1}) and (\ref{spc2}) all the non--trivial information is
contained in the SP level density $\rho (\epsilon)$. We use here the
continuous approximation, in which the discreteness of the SP energy levels
$\epsilon_j$ is ignored and $\rho (\epsilon)$ is replaced by a smooth
function.
We assume moreover that the high energy growth of the integrated density of
states is well approximated, on average, by
\begin{equation} \label{nav}
\nc (\epsilon)\approx \epsilon^\nu \ .
\end{equation}
Here $\epsilon$ is an adimensional energy. To recover dimensional quantities
in the formulas below, all energies must be multiplied by some appropriate
factor $\kappa$. The index $\nu$ is a real positive number that can take
arbitrary values depending on the confining potential. For instance, if the
gas is trapped in a one--dimensional potential whose energy levels $\epsilon_j
= j^s$, then $\nu=1/s$. In contrast, when the confining potential is a
D--dimensional harmonic oscillator, then $\nu = D$, while when it is a
D--dimensional box (hard wall cavity potential), then $\nu=D/2$ (and, for
instance, $\kappa = (V/6\pi^2)^{2/3} (2 m /\hbar^2)$ when $D=3$, where $V$ is
the volume of the cavity and $m$ the mass of the particle).

In the approximation (\ref{nav}) the weight in the sum (\ref{spc2}) of the SP
ground--state is effectively fixed to zero. Under these conditions, $\ec
(\alpha, \beta)$ (and $E$) represent the excitation energy of the gas,
measured with respect to the ground--state energy where all particles are in
the $j=0$ state. For $\nc (\alpha, \beta)$ (and $N$) a problem appears for
$\nu>1$, when condensation may happen. In this case, Eq.~(\ref{spc2}) takes
into account only the thermal cloud. If needed, we will explicitly incorporate
the ground--state occupancies in the calculations.

From Eqs.(\ref{omegadef}) and (\ref{spc2}), using (\ref{nav}) and consequently
$\rho(\epsilon)=\nu \epsilon^{\nu-1}$, the energy and grand potential are
simply related by $\ec (\alpha, \beta) = - \nu \ \Omega (\alpha, \beta)$. The
entropy (\ref{sdef}) may thus be written, taking into account the condition
(\ref{spc2}),
\begin{equation} \label{sdefa}
\sca (\alpha,\beta) = \left( 1 + 1/\nu \right) \beta E - \alpha N \ .
\end{equation}
For any finite $N$, $\alpha$ is easily seen from Eq.~(\ref{spc1}) to be
negative. A standard series expansion of the denominator in Eqs.(\ref{spc1})
and (\ref{spc2}) in terms of $z=\exp (\alpha)$, where $0 < z < 1$, allows to
write, in the continuous approximation, the two saddle point conditions as
\begin{eqnarray} \label{spc3}
\nc (\alpha, \beta) &=& \frac{\Gamma (\nu + 1)}{\beta^\nu} Li_\nu (z) = N \ , \\ 
\ec (\alpha, \beta) &=& \frac{\nu \Gamma (\nu + 1)}{\beta^{\nu+1}} Li_{\nu+1} 
(z) = E \ , \label{spc4}
\end{eqnarray}
where $Li_\nu (z) = \sum_{k=1}^\infty z^k /k^\nu$ is the polylogarithm
function and $\Gamma$ is Euler function.

Eq.~(\ref{spc3}) shows that $\nc (\alpha, \beta)$ is an increasing function of
$z$. Therefore when $N$ increases at a fixed temperature $T=\beta^{-1}$, $z$
needs to increase to satisfy the equality. As $N \rightarrow \infty$, $z
\rightarrow 1$. In that limit, the energy is easy to obtain and we get $\ec
(0, \beta)=\nu \int_0^\infty d\epsilon \epsilon^\nu /(\exp (\beta \epsilon) -
1)=\theta_\nu /\beta^{\nu+1}$, where
\begin{equation} \label{tetanu}
\theta_\nu = \nu \Gamma (\nu +1) \zeta(\nu +1)
\end{equation}
($\zeta (z) = Li_1 (z)$ is the Riemann zeta function). From Eq.~(\ref{spc4}),
we get the following relation between inverse temperature and excitation
energy,
\begin{equation} \label{betanu}
\beta =\beta_{\nu}= [ \theta_\nu /E ]^{1/(1+\nu)} \ .
\end{equation}
Using this expression for $\beta$ and setting $\alpha=0$ in Eq.~(\ref{sdefa}),
we get to leading order in a high energy expansion (i.e., large energies
compared to the spacing between SP energy levels)
\begin{equation} \label{rhoainf}
\rmb (E,\infty) = \exp \left[ (1+1/\nu) (\theta_\nu E^\nu)^{1/(\nu +1)} \right] 
\ .
\end{equation}
For $\nu=1$ this equation reproduces the well known asymptotic result for an
equidistant spectrum $\epsilon_j = j$ (1D harmonic potential), $\rmb
(E,\infty) = {\rm e}^{2 \sqrt{\pi^2 E/6}}$, obtained by Hardy and Ramanujan in
the partition problem \cite{hr} . It was generalized to arbitrary 1-D
potentials $\epsilon_j \propto j^{1/\nu}$ (partitions into non-integral powers
of integers) in \cite{aa}. In the present context, Eq.~(\ref{rhoainf}) is valid
for any system whose average counting function behaves (asymptotically) like
Eq.~(\ref{nav}) (see also Ref.\cite{ht}). For instance, it holds for a 3-D
harmonic potential ($\nu=3$), or a 2-D box of arbitrary shape (or billiard),
($\nu=1$).

Equation (\ref{rhoainf}) describes the density in the limit of an infinite
number of particles for a large but finite excitation energy $E$. In the
opposite limit, of a large excitation energy at a fixed number of particles,
the density behaves quite differently. This is the Maxwell-Boltzmann limit,
where the gas behaves classically. From Eq.~(\ref{spc3}), keeping $N$ fixed and
increasing the temperature (e.g., decreasing $\beta$), it follows that $z
\rightarrow 0$ to satisfy the equality. Then $Li_\nu (z) \approx z$ for any
$\nu$, and the stationary phase conditions (\ref{spc3})--(\ref{spc4}) become
$N=\Gamma (\nu+1) z/\beta^\nu$ and $E= \nu \Gamma (\nu+1) z/\beta^{\nu+1}$.
The relation between temperature and excitation energy now is
\begin{equation} \label{equip}
E=\nu N T \ .
\end{equation}
This simple equation generalizes, to an arbitrary confining potential, the
well-known equipartition of energy valid for quadratic Hamiltonians. It
provides a precise relation between a quantum spectral property (the index
$\nu$) and the partition of energy in the classical limit. From the previous
form of the stationary phase conditions when $z \rightarrow 0$ we also get
$\alpha = \log (\beta^\nu N/ \Gamma (\nu+1))$. Using this relation for
$\alpha$ and Eq.~(\ref{equip}) for $\beta$ in Eq.~(\ref{sdefa}), the many--body
level density now takes the form
\begin{equation} \label{rhomb}
\rmb (E,N) = \left[ \frac{\Gamma (\nu+1)}{\nu^\nu}\frac{E^\nu}{N^{\nu+1}} 
\right]^N
{\rm e}^{(\nu+1) N} \ .
\end{equation}
In contrast to Eq.~(\ref{rhoainf}), in the classical limit the level density
has a power--law dependence on the excitation energy (similar results in some
particular cases were obtained in \cite{wei}, using different methods). When
$E\gg N\gg1$, using Stirling's approximation this equation may be written as
$\rmb (E,N) = \frac{[\Gamma(\nu+1)]^N}{N!} \left( \begin{array}{c} E \\ \nu N
\end{array} \right) $. Under this form, this result coincides for $\nu=1$
with the result obtained in Ref.\cite{vu} for the asymptotic behavior of the
partition of integers with a maximum number of summands (see also
\cite{el}).

So far, we have derived two distinct behaviors of the level density with
excitation energy:  a stretched-exponential behavior in the
quantum-degenerate gas limit, and a power--law behavior in the high
temperature classical limit. In the classical limit, in any typical
configuration of energy $E$ all the particles of the gas are excited, while in
the quantum-degenerate case only a finite fraction of the total number of
particles contribute to the excitation energy (the remaining particles are in
the ground state). To have a better understanding, in the latter case, of the
distribution of the number of excited particles among all the configurations
of energy $E$, and to gain some insight about the transition between the two
extreme regimes, we now compute, starting from the degenerate-gas limit
$z\rightarrow 1$, finite $N$ corrections.

We are interested in particular in computing the relative density $F (E,N) =
\rmb (E,N)/\rmb (E,\infty)$. This quantity gives, among all the possible
states of energy between $E$ and $E + dE$, the fraction of those whose number
of excited particles does not exceed $N$. Interestingly, we find three
distinct behaviors for $F (E,N)$, depending on the value of $\nu$. In terms
of a suitable rescaled variable $x$ that depends on $N$, $E$ and $\nu$ (cf
below), the fraction $F (E,N)$ behaves as
\begin{eqnarray} 
\nu = 1 : F (E,N) &=& \exp(-\exp(-x)) \ ,  \label{lamb1} \\
 0 \! < \! \nu \! < \! 1 :  F (E,N) &=& \left\{ \begin{array}{l} 0 
\;\;\;\;\;\;\;\;\;\;\;\;\;\;\;\;\;\;\;\;\;\;\;\;\;\;\;\, x \leqslant 0 \\ 
\exp[-x^{-\nu/(1-\nu)}] \;\;\;\; x > 0 \;\;\;\;\;\;\;\;  \end{array} \right. 
\label{lamb2} \\
\nu > 1 : F (E,N) &=& \left\{ 
\begin{array}{l} \exp(- |x|^\gamma ) \;\;\;\;\;\;\;\;\;\;\;\; x\leqslant 0 \\
1 \;\;\;\;\;\;\;\;\;\;\;\;\;\;\;\;\;\;\;\;\;\;\;\;\;\;\; x>0 \ .
\end{array} \right. \label{lamb3}
\end{eqnarray}
In the latter case, the index $\gamma$ depends on the precise value of $\nu$
(see Eq.~(\ref{xscal1})). These three distributions are known as Gumbel,
Fr\'echet and Weibull, respectively. They are the three universal limit
distributions well known in the theory of extreme value statistics of
uncorrelated random variables~\cite{Gumbel}. Below we outline the main steps
in the derivation of Eqs.(\ref{lamb1})--(\ref{lamb3}) (details will be
published elsewhere).

To prove Eqs.(\ref{lamb1})--(\ref{lamb3}) one needs to compute from
Eqs.(\ref{spc3}) and (\ref{spc4}) $\alpha (E,N)$ and $\beta (E,N)$, and to
replace them in the expression (\ref{sdefa}). This is done for large but
finite values of the particle number $N$, i.e. in the limit $z={\rm e}^{-\eta}
\rightarrow 1$, where $\eta = - \alpha$ is a small positive parameter.
Mathematically, this requires the computation of the leading order behavior,
when $\eta \rightarrow 0^+$, of the polylogarithmic function $Li_\nu ({\rm
  e}^{-\eta})$. This can be achieved either by a direct computation of the
integrals, or by relying on existing results \cite{flajo}. Once this is done,
$F(E,N)$ is obtained by dividing the density by $\rmb (E,\infty)$ in
Eq.~(\ref{rhoainf}).

{\bf Case I: $\nu = 1$}. We find that the appropriate scaling variable for the
limiting distribution Eq.~(\ref{lamb1}) is
\begin{equation} \label{xscal0}
\hspace{-1.0in} \nu = 1 : \hspace{0.5in}  x=\beta_1 \ N + \log \beta_1 \ ,
\end{equation}
where $\beta_1 = (\pi^2 /6 E)^{1/2}$ was defined in Eq.~(\ref{betanu}). It
follows from Eq.~(\ref{lamb1}) that the asymptotic value for the typical
number of excited bosons for states of energy $E$ is $\beta_1^{-1} \ \log
\beta_1^{-1}$. In the case of an equidistant spectrum $\epsilon_j = j$, this
result reproduces the one obtained by Erd\"os and Lehner in Ref.~\cite{el} for
the partition problem.

{\bf Case II: $0 < \nu < 1$}. From the procedure described above, now we
obtain for $F(E,N)$ the Fr\'echet distribution, Eq.~(\ref{lamb2}), with the
rescaled variable given by
\begin{equation} \label{xscal11} \hspace{-0.9in} 0 < \nu < 1 : \hspace{0.4in}
x=\frac{N}{c_\nu E^{1/(1+\nu)}} \ ,
\end{equation}
where $c_\nu = [(1-\nu)/\nu]^{(1-\nu)/\nu} [\Gamma (1+\nu) \Gamma
(1-\nu)]^{1/\nu} / \theta_{\nu}^{1/(1+\nu)}$. Note that in Eq.~(\ref{lamb2})
the exponent $\nu/(1-\nu)$ is positive in the corresponding range of $\nu$.
This distribution implies that the typical number of excited bosons for states
of energy $E$ is $c_\nu E^{1/(1+\nu)} / 2$. However, note that the
distribution is strongly asymmetric, with a power-law decay (toward 1) for $N$
much larger than the typical value.

{\bf Case III: $\nu > 1$}. This case is slightly more complicated than the
previous ones, because of the presence of a phase transition. In contrast with
the previous cases, as $N$ increases and $z \rightarrow 1$ in Eq.~(\ref{spc3})
at fixed $\beta$, the function $Li_\nu (z)$ tends to a finite value. At
constant temperature, there is thus a critical number $N_c = \Gamma (1+\nu)
\zeta(\nu)/\beta^{\nu}$ of bosons that can be hosted by the thermal cloud,
above which a Bose-Einstein condensation starts. We find that the relevant
variable in this case is not $N$ but the difference $N - N_c$. The behavior of
the distribution is different according to whether $N$ is smaller or larger
than $N_c$. When $N \leqslant N_c$, the exponent $\gamma$ and the rescaled
variable $x$ in Eq.~(\ref{lamb3}) depend on the precise value of $\nu$. Three
different regimes are found, summarized as follows
\begin{eqnarray} 
1 \! < \! \nu \! < \! 2 : \gamma = \frac{\nu}{\nu-1} ; \ 
x = \frac{\beta_\nu (N-N_c)}{[\nu \Gamma (\nu-1) \Gamma (2-\nu)]^{1/\nu}}
\;\;\;\;\;\;\;\;\; && \label{xscal1} \\
\nu = 2 : \gamma = 2 ; \ x= \left\{ \frac{\beta_\nu^\nu /\nu}
{\log[(\beta_\nu^\nu (N_c-N) /\nu)]} \right\}^{1/2} \hspace{-0.2in} (N-N_c)
\;\;\; && \label{xscal2} \\
\nu > 2 : \gamma = 2 ; \ x= \frac{\beta_\nu^{\nu/2} (N-N_c)}{[\Gamma(\nu+1) 
\zeta(\nu-1)]^{1/2}} \hspace{0.87in}  && \label{xscal3}
\end{eqnarray}
where $\beta_{\nu}$ is given in Eq.~(\ref{betanu}). Finally, for any $\nu>1$
and $N>N_c$ (that corresponds to $x>0$), a macroscopic fraction of the
particles is in the ground state. These particles do not contribute to the
excitation energy, and their precise number is unimportant. The behavior of
the system is thus identical to that of the $N\rightarrow \infty$ limit,
implying $F (E,N) = 1$ for $N>N_c$ (or $x>0$). This completes the
demonstration of the Weibull distribution, Eq.~(\ref{lamb3}).

The connection to the number partitioning problem becomes evident if one
chooses $\epsilon_j=j$ and $E$ to be a positive integer. The relation
$E=\sum_{j=1}^{\infty} n_j j$ then corresponds to partitioning $E$ into
non-zero integers and $N_{\rm ex}=\sum_{j=1}^{\infty} n_j$ corresponds to the
number of terms or summands in a given configuration of partition. The ratio
$F(E,N)=\rmb (E,N)/\rmb (E,\infty)$ then represents the probability that the
number of summands in a random partition of integer $E$ is less than or equal
to $N$. The corresponding limiting Gumbel law for $F(E,N)$ was first proved by
Erd\"os and Lehner by rigorous methods~\cite{el}. Our results provide a
generalization of this theorem to an arbitrary set of summands characterized
by the growth law Eq.~(\ref{nav}). The particular case $\epsilon_j= j^s$ with
$s> 0$ corresponds to partitioning an integer $E$ into sums of $s$-th powers
of non-zero integers. For example, for $s=2$, the integer $5$ can be
partitioned into sums of squares as $5=2^2+1^2=1^2+1^2+1^2+1^2+1^2$. We have
shown that while for $s=1$ we recover the Gumbel law, the limiting
distribution of $F(E,N)$ is Fr\'echet for $s>1$ (or $0< \nu<1$) and Weibull
for $s<1$ (or $\nu>1$).

In conclusion, we have shown that the density of states of a system of
independent bosons is described in a suitable scaling limit by the three
limiting laws of extreme value theory. This result has a universal character
since it depends only on a single parameter $\nu$ that governs the large
energy asymptotic average behavior of the SP energy spectrum (and is
independent, for instance, of the fluctuation properties of the SP spectrum).
The derivative of the fraction $F(E,A)$ is related to the probability density
of the number of excited particles. Our general results should allow to
recover the moments of the distribution of the ground state occupation numbers
computed in, e.g., Refs.~\cite{WH}. A probabilistic interpretation of our
results in the light of~\cite {Vershik} may shed further light on the
connection with extreme value theory.

This work was supported by grants ACI Nanoscience 201, ANR NT05-2-42103,
ANR-05-Nano-008-02 and the IFRAF Institute.


\end{document}